\documentclass[aps,prb,twocolumn,showpacs,superscriptaddress]{revtex4}

\usepackage{graphicx}
\usepackage{amsmath}
\graphicspath{{fig/}{figsgaoerb/}}
\usepackage[colorlinks,linkcolor=blue,anchorcolor=blue,citecolor=blue]{hyperref}

\begin{document}
\title{Single crystal growth and superconductivity in RbNi$_2$Se$_2$}	
\author{Hui Liu}
\author{Xunwu Hu}
\affiliation{Center for Neutron Science and Technology, Guangdong Provincial Key Laboratory of Magnetoelectric Physics and Devices, School of Physics, Sun Yat-Sen University, Guangzhou, 510275, China }
\author{Hanjie Guo}
\affiliation{Neutron Science Platform, Songshan Lake Materials Laboratory, Dongguan, Guangdong, 523808, China}
\author{Xiao-Kun Teng}
\affiliation{Department of Physics and Astronomy, Rice University, Houston, TX 77005, USA}
\author{Huanpeng Bu}
\affiliation{Neutron Science Platform, Songshan Lake Materials Laboratory, Dongguan, Guangdong, 523808, China}
\author{Zhihui Luo}
\author{Lisi Li}
\author{Zengjia Liu}
\author{Mengwu Huo}
\author{Feixiang Liang}
\author{Hualei Sun}
\author{Bing Shen}
\affiliation{Center for Neutron Science and Technology, Guangdong Provincial Key Laboratory of Magnetoelectric Physics and Devices, School of Physics, Sun Yat-Sen University, Guangzhou, 510275, China }
\author{Pengcheng Dai}
\affiliation{Department of Physics and Astronomy, Rice University, Houston, TX 77005, USA}
\author{Robert J. Birgeneau}
\affiliation{Department of Physics, University of California, Berkeley, California 94720, USA }
\affiliation{Materials Science Division, Lawrence Berkeley National Laboratory, Berkeley, California 94720, USA}
\author{Dao-Xin Yao}
\affiliation{Center for Neutron Science and Technology, Guangdong Provincial Key Laboratory of Magnetoelectric Physics and Devices, School of Physics, Sun Yat-Sen University, Guangzhou, 510275, China }
\author{Ming Yi}
\affiliation{Department of Physics and Astronomy, Rice University, Houston, TX 77005, USA}
\author{Meng Wang}
\email{wangmeng5@mail.sysu.edu.cn}
\affiliation{Center for Neutron Science and Technology, Guangdong Provincial Key Laboratory of Magnetoelectric Physics and Devices, School of Physics, Sun Yat-Sen University, Guangzhou, 510275, China }
	
	\begin{abstract}
We report the synthesis and characterization of RbNi$_2$Se$_2$, an analog of the iron chalcogenide superconductor Rb$_x$Fe$_2$Se$_2$, via transport, angle resolved photoemission spectroscopy, and density functional theory calculations. A superconducting transition at $T_{c}$ = 1.20 K is identified. In normal state, RbNi$_2$Se$_2$ shows paramagnetic and Fermi liquid behaviors. A large Sommerfeld coefficient yields a heavy effective electron mass of $m^{*}\approx6m_{e}$. In the superconducting state, zero-field electronic specific-heat data $C_{es}$ can be described by a two-gap BCS model, indicating that RbNi$_2$Se$_2$ is a multi-gap superconductor. Our density functional theory calculations and angle resolved photoemission spectroscopy measurements demonstrate that RbNi$_2$Se$_2$ exhibits relatively weak correlations and multi-band characteristics, consistent with the multi-gap superconductivity.

	\end{abstract}
	
	\maketitle
	
	\section{Introduction}	

Since the discovery of copper oxide superconductors, researchers have extensively searched for superconductivity in materials with transition metals\cite{Bednorz1986,Takada2003}. Significant progress has been made in iron pnictide and chalcogenide compounds, where several structural systems have been identified with the highest $T_c$ of 55 K achieved in LaFeAsO\cite{Ren2008}. Superconductivity was also observed in the compounds consisting of Cr and Mn under pressure, such as CrAs, KCrAs, and MnP\cite{Wu2014,Mu2017,Cheng2015}. Among all of them, superconductivity has been found to be in the vicinity of an antiferromagnetic (AF) order, suggesting that spin fluctuations may play an important role in the mechanism of superconductivity. Nickel oxide materials have analogous structures with copper oxide superconductors. Superconductivity with $T_c=9-15$ K has also been observed in films of nickel-based compounds\cite{Li2019a,Gu2022}.

$A$Ni$_2$Se$_2$ ($A=$ K, Cs, and Tl) crystalizes in the ThCr$_2$Si$_2$ structure and shows metallic behavior and Pauli paramagnetism. At low temperatures, superconductivity emerges with the superconducting (SC) transition temperatures of $T_{c}$ $\approx$ 0.8 K for KNi$_2$Se$_2$ polycrystals, 2.7 K for CsNi$_2$Se$_2$, and 3.7 K for TlNi$_2$Se$_2$ single crystals\cite{Neilson2012,Chen2016,Wang2013}. While K$_{0.95}$Ni$_{1.86}$Se$_2$ single crystals do not show superconductivity down to 0.3 K, yielding that the superconductivity is sensitive to the stoichiometry of the samples\cite{Lei2014}. As a comparison, \textit{A}$_x$Fe$_{2}$Se$_2$ system exhibits $T_c$s ranging from 20 $-$ 30 K. With different amount of iron vacancies, \textit{A}$_x$Fe$_{2-\delta}$Se$_2$ exhibits a variety of AF orders and iron vacancy orders\cite{Guo2010,Fang2013,Dai2015,Wang2016}. The replacement of Fe by Co suppresses the superconductivity and induces a ferromagnetic (FM) order in RbCo$_2$Se$_2$\cite{Yang2013,Huang2021}. The $A$Ni$_2$Se$_2$ superconductors with a formal valence of Ni$^{1.5+}$ have been revealed to exhibit remarkable properties. In particular, they usually exhibit a large Sommerfeld coefficient $\gamma$, suggesting a large density of states and unconventional pairing at low temperature\cite{Wang2013,Chen2016}. One possibility that was proposed was that the large Sommerfeld coefficient might be induced by local charge order\cite{Neilson2012}. However, angle resolved photoemission spectroscopy (ARPES) measurements yield weak electronic correlations in KNi$_2$Se$_2$ and the origin of the large Sommerfeld coefficient may be driven by the large density states and the Van Hove singularity in the vicinity of the Fermi energy\cite{Fan2015}. 

In this work, we report the successful synthesis and characterization of RbNi$_{2}$Se$_{2}$ single crystals. The crystal structure, electronic band structure, and transport properties have been investigated. We find that RbNi$_{2}$Se$_{2}$ is a Pauli paramagnetism and exhibits a SC transition at $T_{c}$ = 1.20 K. Normal-state specific heat measurements suggest an effective electronic mass enhancement with $m^{*}\approx6m_{e}$. In SC state, a two-gap BCS model can match well with the zero-field electronic specific heat, indicating that RbNi$_{2}$Se$_{2}$ is a multi-gap superconductor. Comparison with the density functional theory (DFT) calculations and ARPES measurements reveals that RbNi$_{2}$Se$_{2}$ is a weakly correlated superconductor with multi bands crossing the Fermi level.

	\begin{figure}[t]
	\centering
	\includegraphics[width=7cm]{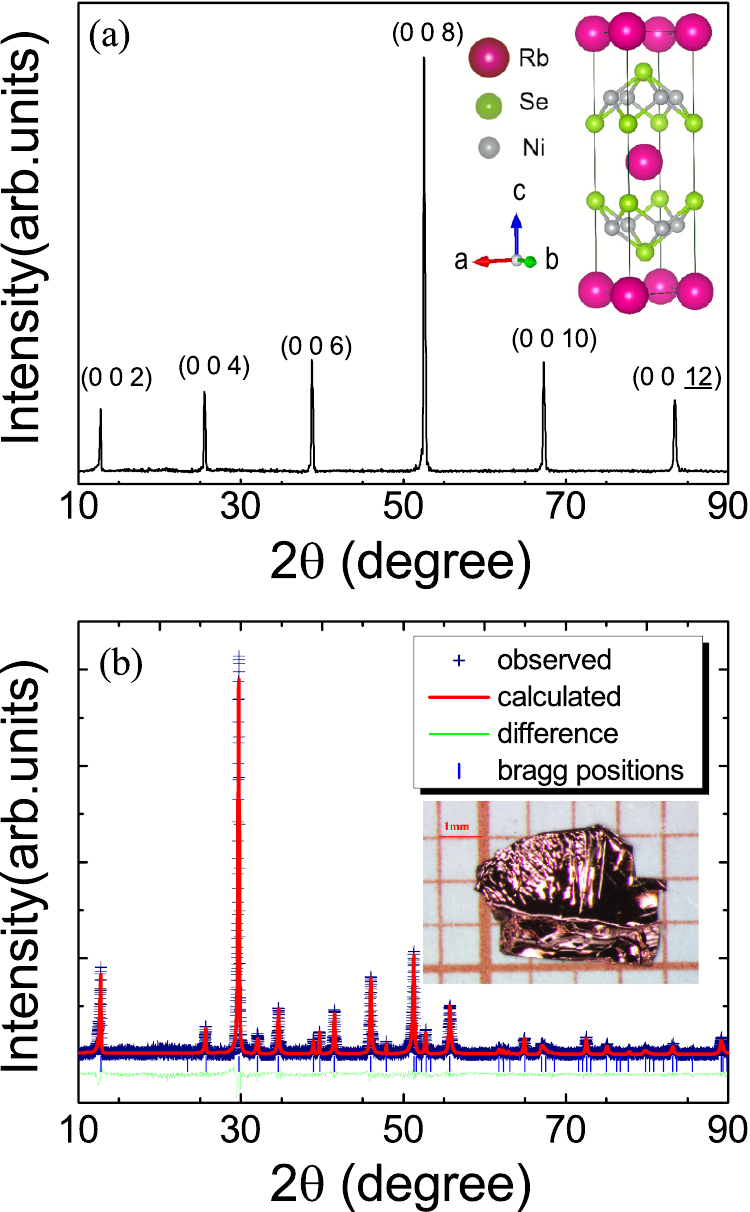}
	\caption{(a) XRD measurement on the $ab$ plane of RbNi$_2$Se$_{2}$ single crystal. The inset shows the crystal structure. The pink, green, and gray balls represent Rb, Se, and Ni ions. (b) A pattern of XRD measured on powder samples. The inset is a photo of RbNi$_{2}$Se$_{2}$ single crystal.}		
	\label{fig1}
	\end{figure}

	\section{Experimental and calculation details}
	
Single crystals of RbNi$_2$Se$_2$ were grown by the self-flux method. First, the precursor NiSe was prepared by heating Ni powders and Se pellets at $500^\circ$C. Then, NiSe powders and Rb were put into alumina crucibles according to stoichiometry and sealed in an evacuated silica tube. The mixture was kept at $150^\circ$C for 5 h, then heated to $760^\circ$C in 40 h and kept for 5 h, after that heated to $1050^\circ$C in 40 h and held for 5 h. Finally, the temperature was cooled down to $700^\circ$C at a rate of $3.5^\circ$C/h. To prevent the reaction of Rb with water and air, all of the processes were conducted in an argon-filled glove box. Shiny plate-like single crystals with a typical size of 4 $\times$ 5 $\times$ 1 mm$^{3}$ were grown as shown in the inset of Fig. \ref{fig1}(b). 
 
\begin{table}
	\caption{Single-crystal of RbNi$_2$Se$_2$ refinement at 150 K.}
	\setlength{\tabcolsep}{1mm}
	\begin{tabular}{c|c}
		\hline \hline 
		Formula weight			&223.14	 \\
		Crystal system 			&Tetragonal	\\
		Space group 			&I4/mmm\\
		Unit-cell parameters	        &a = b = 3.9272 (3) \AA\      \\ 
		                                          &c = 13.8650(5) \AA\          \\
		                                          &$\alpha=\beta =\gamma = 90^\circ$\\
		Atomic parameters                                              \\
		Rb						&2b(0,0,1/2)\\
		Ni						&4d(0,1/2,3/4)\\
		Se						&4e(1/2,1/2,0.6502(1))\\
		Density					&3.466 g/cm$ ^{3} $\\
		F(000)					&198\\
		Radiation				        &Mo $\textit{K}\alpha$ ($\lambda$ = 0.7107\AA)\\
		2$\theta$ for data collection	&$10.794^\circ$ to $60.562^\circ$ \\
		Index ranges			&-4 $ \leq $ h $ \leq $ 5, -5 $ \leq $ k $ \leq $ 5,\\ &-11 $ \leq $ l $ \leq $ 18\\
		Reflections collected	        &808\\
		Independent reflections	&119\\
		Data/restraints/parameters		&119/0/5\\
		Goodness-of-fit on F$ ^{2} $			&1.204\\
		Final R indexes [I $\geq$ 2$\sigma$(I)]			&$ \textit{R}_{1} $ = 0.0378, $ \textit{wR}_{2} $ =0.1001\\
		Largest diff. peak/hole/e \AA$^{-3}$					&3.06/-2.04	\\  \hline\hline
	\end{tabular}
	\label{table1}
\end{table}

Single crystal x-ray diffraction (XRD) were conducted on a SuperNova (Rigaku) x-ray diffractometer. The sample was blowed by N$_2$ during the data collection to avoid exposure to air. The elemental analysis was measured by using an energy-dispersive x-ray spectroscopy (EDS) (EVO, Zeiss). Electrical transport,  magnetic and specific heat measurements were performed on a physical property measurement system (PPMS, Quantum Design). The in-plane resistivity $\rho_{ab}(T)$ was measured using the standard four-probe method on a rectangular sheet crystal to keep current flowing in the \textit{ab}-plane. The Vienna \textit{Ab initio} Simulation Package (VASP) was employed for the DFT calculations\cite{Kresse1993}. ARPES measurements were performed on a helium-lamp based system with a DA30 electron analyzer. Single crystals were cleaved \emph{in-situ} in ultra-high vacuum with a base pressure better than $5\times10^{-11}$ Torr at 30 K. Energy and angular resolutions were better than 20 meV and $0.1^\circ$, respectively.

	\section{Results and discussions}

All peaks from single crystal XRD can be indexed with the ThCr$_2$Si$_2$-type structure (space group: I4/mmm), which is illustrated in the inset of Fig. \ref{fig1}(a). The determined lattice parameters are $a=b=3.9272 (3)$, and $\textit{c} = 13.8650(5)$ \AA\ at 150 K with the volume of unit cell between that of KNi$_2$Se$_2$ and CsNi$_2$Se$_2$. Details of the atom coordinates and other key information are shown in Table \ref{table1}.
To show the quality of the samples, we present a $\theta$$-$2$\theta$ scan of a single crystal along the $(H=0, K=0, L)$ in Fig. \ref{fig1}(a) and an XRD pattern on a powder sample in Fig. \ref{fig1}(b), where $(H, K, L)$ are Miller indices in reciprocal lattice units. No peaks from impurity could be identified. The EDS results for several single crystals are rather homogenous and the determined average atomic ratios are Rb:Ni:Se = 1.16(4):2.04(3):2.00(6) when the content of Se is normalized to be 2, close to the stoichiometry of RbNi$_2$Se$_2$.

\begin{figure}[t]
	\centering
	\includegraphics[width=7cm]{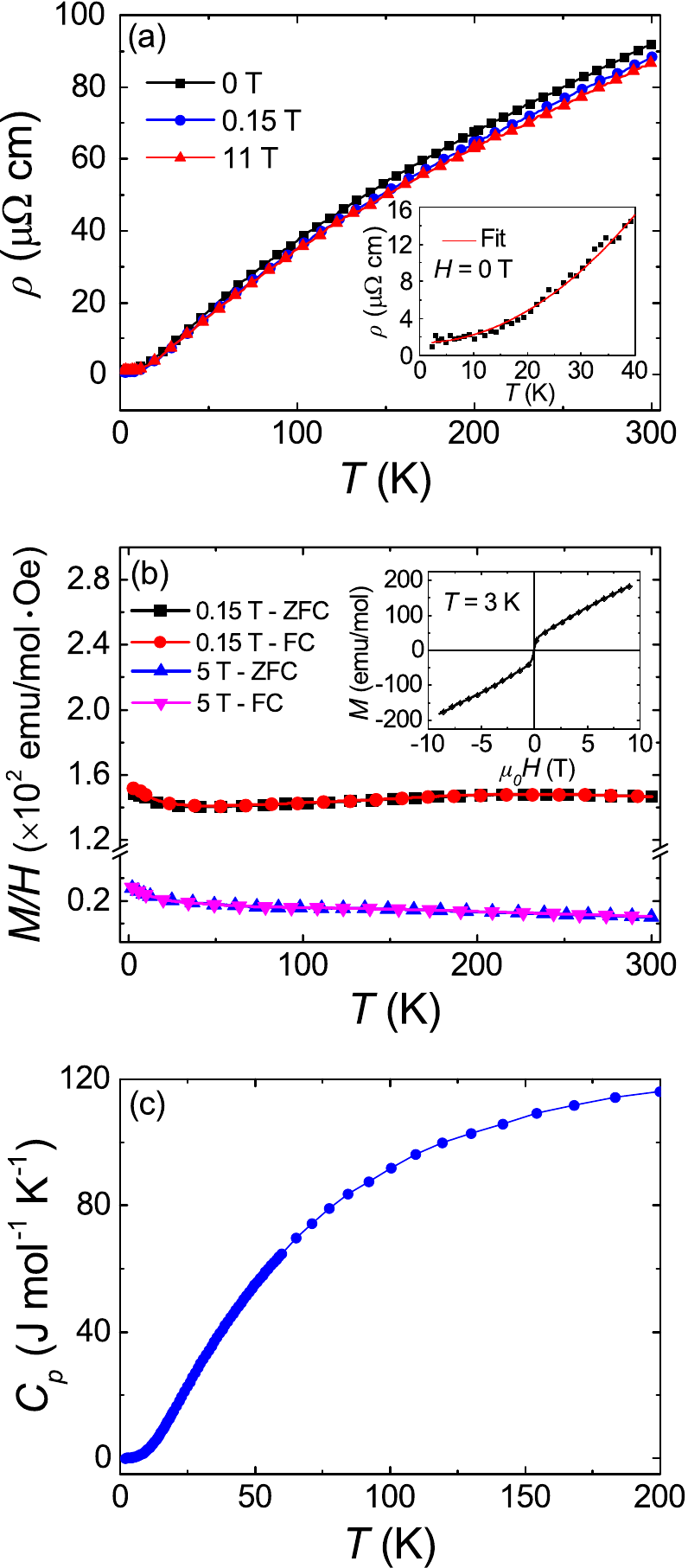}
	\caption{(a) Temperature dependence of in-plane resistivity $\rho_{ab}(T)$ with magnetic fields \textit{H} = 0, 0.15, and 11 T along the \textit{c} axis. The inset shows the fitted result using $\rho_{ab}($\textit{T}$ )=\rho_{0}+A$\textit{T}$ ^{2}$ from 1.8 to 40 K. The red line is a fitting curve. (b) Temperature dependence of ZFC and FC magnetic susceptibility with applying various external magnetic fields along the $c$ axis. The inset shows a magnetization hysteresis loop at 3K. (c) Temperature dependence of $\textit{C$ _{p} $}$($\textit{T}$ ) from 2 to 300 K.}
	\label{fig2}
\end{figure}

Temperature dependence of the resistivity for RbNi$_2$Se$_2$ is shown in Fig. \ref{fig2}(a). The electric current is applied in the \textit{ab}-plane. The value of $\rho_{ab}$ is about 92.1 $\mu\Omega\cdot$cm at 300 K and only about 1.5 $\mu\Omega\cdot$cm at 1.8 K. The residual resistivity ratio (\textit{RRR}) of 61.4 [$\rho_{ab}(300\ \text{K})/\rho_{ab}(1.8\ \text{K})$] suggests remarkable metallicity and high quality of the single crystals\cite{Lei2014,Ehrlich1968,Bohmer2016a}. The resistivity measured at 0, 0.15, and 11 T shows a metallic behavior without any anomaly or strong magnetic field dependence. The resistivity $\rho_{ab}$ below 40 K can be well described by the equation $\rho_{ab}($\textit{T}$ )=\rho_{0}+A$\textit{T}$ ^{2}$ as shown in the inset of Fig. \ref{fig2}(a), where $\rho_{0}$ = 1.211 $\mu\Omega\cdot$cm and \textit{A} = 0.012 $\mu\Omega\cdot$cm/K$^{2}$, revealing a paramagnetic Fermi liquid behavior\cite{Analytis2014}. The magnetic susceptibility is nearly independent of temperature, yielding a Pauli paramagnetic behaviour as shown in Fig. \ref{fig2}(b). A weak FM sign is revealed from the hysteresis loop shown in the inset of Fig. \ref{fig2}(b), which could be ascribed to a small amount of Ni impurity\cite{Li2020}. The specific heat from 2 to 200 K shown in Fig. \ref{fig2}(c) also suggests that no phase transition occurs in this temperature range.

\begin{figure*}[t]
	\centering
	\includegraphics[width=15cm]{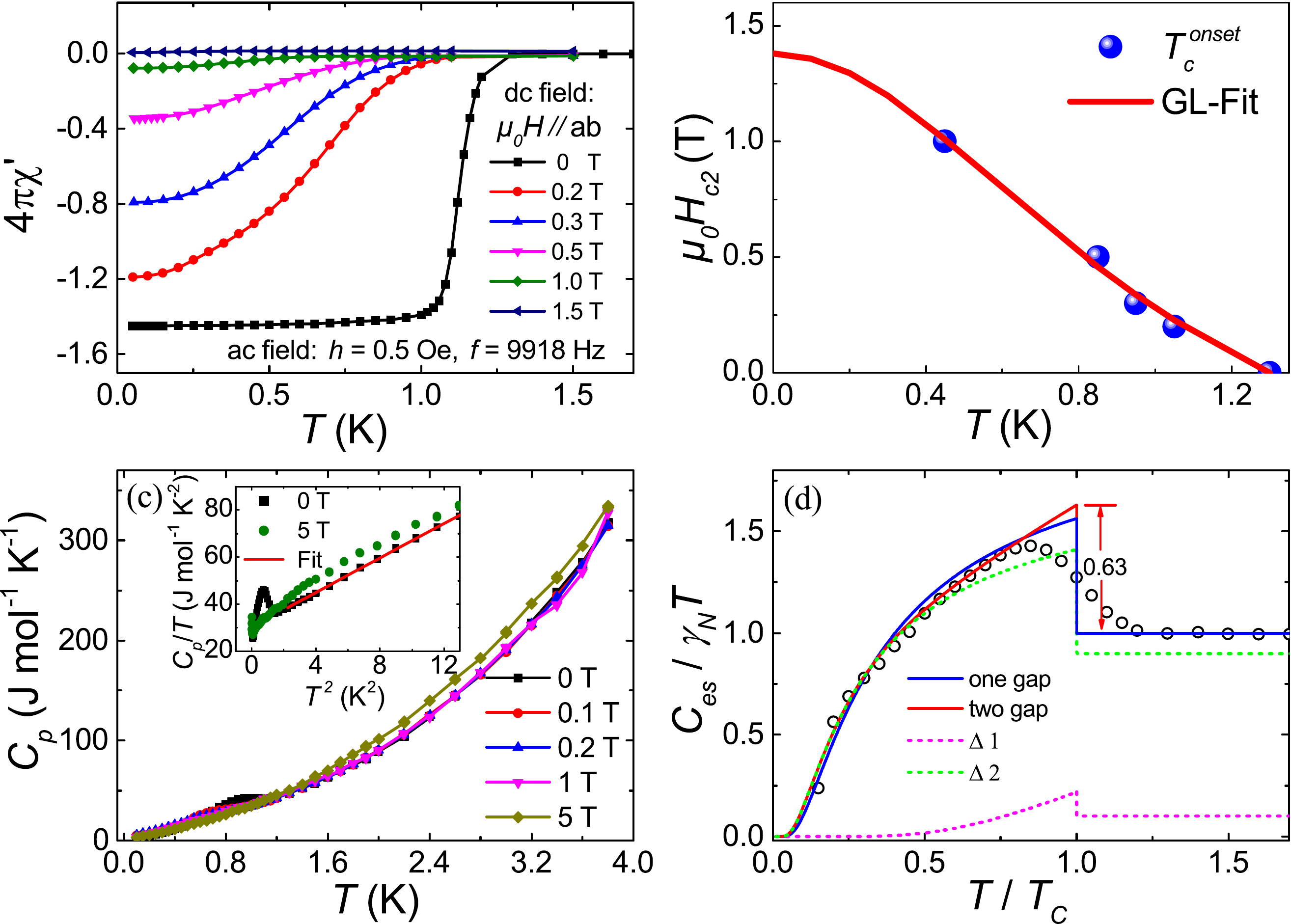}
	\caption{(a) Diamagnetization of superconductivity under various magnetic fields. (b) Upper critical field $H_{c2}$, as a function of temperature. The red solid line shows a fit with the Ginzburg-Landau expression as discussed in the text. (c) The low-temperature specific heat of RbNi$_2$Se$_2$, measured at various fields near superconducting transition. The inset shows the Schotkky anomaly at zero field and 5 T and a fitting of $\textit{C}_{p}$ in normal state. (d) Reduced temperature $\textit{T}/\textit{T}_{c}$ dependence of electronic specific heat divided by temperature and $\gamma _{N} $, $ \textit{C}_{es} / \gamma _{N} \textit{T} $, in the SC state at zero field, where $\textit{C}_{es} = C_{p} - C_{lattice} - C_{N}$. The two solid lines show the fitting curves of the one-gap BCS model and the two-gap model to $ \textit{C}_{es} / \gamma _{N} \textit{T} $, respectively. The dashed lines show the contributions from two different gaps, $\Delta_1$ and $\Delta_2$, respectively. }	
	\label{fig3}
\end{figure*}

To explore the possible superconductivity, we show measurements down to 50 mK in Fig. \ref{fig3}. 
The magnetic susceptibility at low temperatures is shown in Fig. \ref{fig3}(a). A clear diamagnetic response appears below 1.20 K under zero field, indicating a SC transition. The transition shifts to lower temperatures with an increase of the $dc$ bias field, consistent with the Meissner effect of superconductivity. With knowing $T_{c}$ $\approx$ 0.8 K for KNi$_2$Se$_2$\cite{Neilson2012} and $T_c\approx$ 2.7 K for CsNi$_2$Se$_2$\cite{Chen2016}, the diamagnetic response at 1.20 K should correspond to the SC transition of RbNi$_2$Se$_2$. We plot the onset SC transition temperatures at various magnetic fields and fit the upper critical field $\mu_{0}H_{c2}(0)$ using the Ginzburg-Landau theory with the formula $\mu_{0}H_{c2}(T) = \mu_{0}H_{c2}(0) \times (1 - t^{2})/(1 + t^{2})$, where $t$ is the reduced temperature $t = T / T_{c}$. As shown in Fig. \ref{fig3}(b), the resultant $\mu_{0}H_{c2}(0)$ = 1.38 T is within the Pauli limit,  $\mu_{0}H^{p}_{c2}(0)$ = 2.37 T, indicating a weak coupling behavior\cite{Bao2015a}.

stop here

Figure \ref{fig3}(c) displays the specific heat measured at low temperatures. In normal state, a fit to the specific heat $C_{p}$ from 1.5 to 3.8 K using $C_{p} /T$ = $ \gamma_{N} $+$ \beta T^{2} $ results in the Sommerfeld coefficient $\gamma_{N}$ = 30.30 mJ$\cdot$mol$^{-1}$K$^{-2}$ and $\beta$ = 3.67 mJ$\cdot$mol$^{-1}$K$^{-4}$, as shown in the inset. The Debye temperature $\Theta_{D}$ is estimated to be of 167 K from the equation $\Theta_{D}$ = $(12\pi^{4}NR/5\beta)^{1/3}$, where $N=5$ is the atomic number in each formula unit and $R$ is the ideal gas constant. The $ \gamma_{N} $ for RbNi$ _{2} $Se$ _{2} $ is comparable with that of KNi$ _{2} $Se$ _{2} $ ($\sim$44 mJ$\cdot$mol$^{-1}$K$^{-2}$)\cite{Neilson2012}, CsNi$ _{2} $Se$ _{2} $ ($\sim$77 mJ$\cdot$mol$^{-1}$K$^{-2}$)\cite{Chen2016}, and TlNi$ _{2} $Se$ _{2} $ ($\sim$40 mJ$\cdot$mol$^{-1}$K$^{-2}$)\cite{Wang2013}.The effective mass of electrons $m^{*}$ can be estimated through Eq. \ref{eq1}\cite{DeLong1985}:
\begin{equation}	
{m^{*}} = {\hbar^{2}k_{F}^{2}\gamma_{N}} / {\pi^{2}nk_{B}^{2}}
\label{eq1}
\end{equation}
where $k_{B}$ is the Boltzmann constant and the carrier density $n$ is calculated by the number of electrons ($Z$) per cell volume ($V$). Using a spherical Fermi surface approximation, the Fermi wave vector can be estimated by $k_{F}$ = $(3\pi^{2}n)^{1/3}$. Assuming that Ni contributes 1.5 electrons ($Z$ = 6), we obtain $k_{F}$ = $9.4 \times 10^{9}$ m$^{-1}$. The estimated effective mass of electrons $ m^{*}/m_{e} $ = 6 for RbNi$ _{2} $Se$ _{2}$ is significantly enhanced compared with the bare electron mass $ m_e $. Combining the parameters from fitting the electronic specific heat and the quadratic temperature dependent regime of resistivity, the Kadowaki-Woods ratio, $A$/$\gamma_N^{2}$, is calculated to be 0.94 $\times$ $10^{-5}\mu\Omega\cdot$cm(mol$\cdot$K$^{2}$mJ)$^{2}$, close to $\sim$10$^{-5}\mu\Omega\cdot$cm(mol$\cdot$K$^{2}$mJ)$^{2}$ of heavy fermion systems\cite{Kadowaki1986}. This scaling relation yields RbNi$ _{2} $Se$ _{2} $ with the heavy-fermion behavior.

In SC state, the specific heat data reveals a clear $\lambda$-anomaly under zero field with a maximum at 0.94 K [Fig. \ref{fig3}(c)], suggesting bulk superconductivity. Applying an external magnetic field, the SC transition moves quickly to lower temperatures. As shown in the inset of Fig. \ref{fig3}(c), an upturn appears on $C_p/T$ below $T^2<0.1$ K$^2$. The small upturn could be described by the Schottky anomaly for paramagnetic impurity spins\cite{Neilson2012}, which can be well fitted by $C_{N}$ = $D(H)T^{-2}$ with $D(H=0)$ = 0.07 mJ$\cdot$K$\cdot$mol$^{-1}$.

In order to get information about the SC gap, we extract the electronic specific heat $C_{es}$ by subtracting the phonon contribution $C_{lattice}$ and Schottky anomaly $C_{N}$ from the total $C_{p}$, $C_{es}=C_{p}-C_{lattice}-C_{N}$. The electronic specific heat data  $ \textit{C}_{es}(\gamma _{N}\textit{T})^{-1}$ against $T/T_c$ are shown in Fig. \ref{fig3}(d), where $\gamma _{N}$ has been revealed in the normal state. The $T_{c}$ is determined to be 0.94 K using an equal-area entropy construction. The heat jump at transition is 0.63, smaller than that of the theoretical value of 1.43 in the BCS weak-coupling scenario\cite{Daams1981}. We employ both one-gap and two-gap BCS models $C_{es} =\sum C_{i} exp(-\Delta_i/k_BT$) to fit the electronic specific heat, where $\Delta_{i}$ is the size of the $i^{th}$ SC gap at 0 K. The one-gap model reveals $\Delta_{0}/k_{B}T_{c}$ = 0.30 that is smaller than the value of 1.76 in the BCS theory\cite{Bardeen1957}. In the two-gap model, the total specific heat can be considered as the sum of electronic contributions from two bands. Our fitting yields the sizes of two gaps of $\Delta_{1}/k_{B}T_{c}$ = 2.51 and $\Delta_{2}/k_{B}T_{c}$ = 0.25, respectively. The ratio of the contributions from the two gaps is $\sim1.5:1$ as  presented by the dashed lines in Fig. \ref{fig3}(d). The better fitting using the two-gap model indicates that RbNi$_2$Se$_2$ may be a multi-gap superconductor. 

\begin{figure*}[t]
	\centering
	\includegraphics[width=15cm]{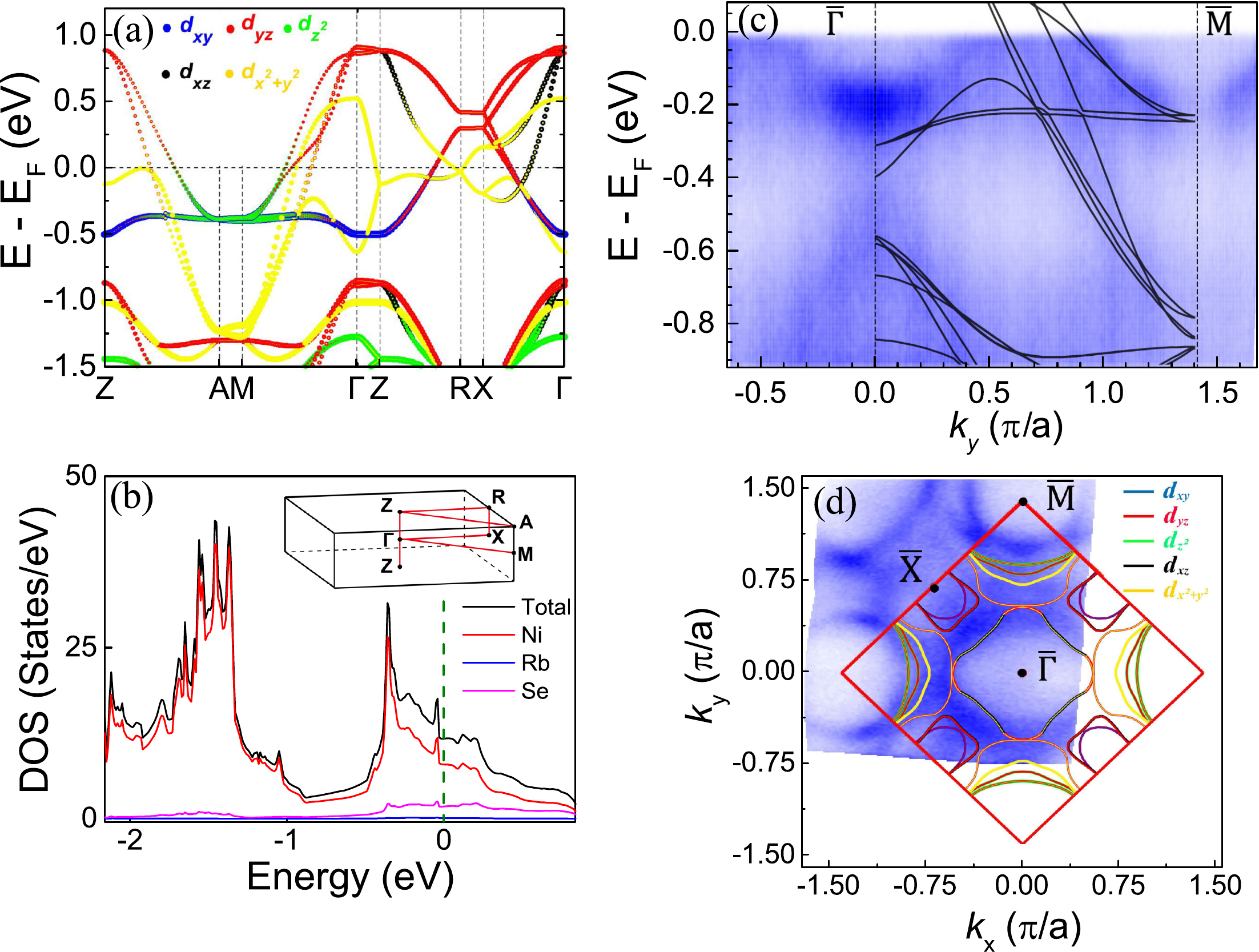}
	\caption{(a) Projected band structure of Ni $3d$ in RbNi$ _{2} $Se$ _{2} $. Colors of the circle represent the different nickel orbitals, and the weight of each orbital is displayed by the size of the circles. The dashed lines are guide to eyes. (b) Density of states (DOS) near the Fermi level of RbNi$ _{2} $Se$ _{2}$. The red, blue and purple curves represent contributions from Ni, Rb and Se, respectively. Inset: The coordinates of the high symmetry $\textbf{k}\textendash$path in reciprocal space of the tetragonal unit cell. (c) Measured spectral images along the high symmetry direction $\varGamma \sim M$. The black lines are band structure calculated by the DFT.  (d) The Fermi surface mapping at 35 K with the Brillouin zone (BZ) marked by a red square. }	
	\label{fig4}
\end{figure*}

The electron-phonon coupling strength is also calculated by employing the inverted McMillan formula\cite{McMillan1968}:
\begin{equation}	
	{\lambda_{ep}} = \frac{1.04+\mu^{*}{\rm ln}(\Theta_{D}/1.45T_{c})}{(1-0.62\mu^{*}){\rm ln}(\Theta_{D}/1.45T_{c})-1.04}
	\label{eq2}
\end{equation}
where $\mu^{*}$ represents the Coulomb repulsion pseudopotential, which we adopt $\mu^{*}$ = 0.13 for this system\cite{Amon2018}. Generally, the $\lambda_{ep}$ for strongly coupled superconductors are close to 1, and $\lambda_{ep}\rightarrow$ 0.5 is viewed as weak coupled superconductors\cite{McMillan1968}. The $\lambda_{ep}$ for RbNi$ _{2} $Se$ _{2}$ is 0.49, suggesting that RbNi$ _{2} $Se$ _{2}$ is a weakly coupled superconductor, consistent with the specific heat analysis.

To check the multi-band character, we conducted DFT calculations and ARPES measurements on the electronic band structure. Figures \ref{fig4}(a) and \ref{fig4}(b) show the calculated electronic bands of Ni ions and the integrated density of states of Ni, Rb, and Se. The electronic states near the Fermi surface are governed by the $3d$ orbitals of Ni ions. In Fig \ref{fig4}(c), the measured electronic structure of RbNi$ _{2} $Se$ _{2}$ is compared with the band structure obtained from the DFT calculations. The calculated curves (black lines) are scaled and overlaid onto the photoemission intensity along the $\Gamma-M$ direction. The theoretical and experimental data match qualitatively with a renormalization factor of about 1.8. This factor is relatively small compared to other iron-based superconductors\cite{yi2017}, indicating that the electronic correlations are moderate. The moderate correlations suggest that the superconductivity is most likely originated from the electron-phonon coupling, similar to the two-gap superconductor MgB$_{2}$\cite{Bouquet2001,Bouquet2001a}. Figure \ref{fig4}(d) shows the photoemission intensity map of the Fermi surface, overlaid with the DFT calculated Fermi surfaces. The multiple Fermi pockets observed are consistent with the multi-band behavior of this compound, reinforcing the condition for the two-gap feature in the SC state. The band structure is reminiscent of the closely-related compound RbCo$_2$Se$_2$\cite{Huang2021}. However, the $3d_{x^2+y^2}$ flat band that induces itinerant ferromagnetism in RbCo$_2$Se$_2$ is observed here to be well below the Fermi level, due to the electron doping resulted from the replacement of Co by Ni.

As $A$Ni$_2$Se$_2$ ($A=$ K, Cs, and Tl) superconductors, RbNi$ _{2} $Se$ _{2}$ also exhibits an enhanced effective electron mass with a large Sommerfeld coefficient from specific heat. However, ARPES data reveal that the electronic correlation is not strong. To understand the reason of the large Sommerfeld coefficient, $\gamma$ is estimated with $\lambda_{ep}$ and $N$(0) at the Fermi level through the relationship\cite{Xiao2021}:
 \begin{equation}	
 	{\gamma} =  \frac{\pi^{2}k_{B}^{2}N(E_{F})(1+\lambda_{ep})}{3}
 	\label{eq3}
 \end{equation}
From the DOS in Fig. \ref{fig4}(b), {$N(E_{F})$ is estimated to be 11.68 states/eV per formula, resulting in $\gamma$ = 40.98 mJ$\cdot$mol$^{-1}$K$^{-2}$ that is close to the experimental value of 30.30 mJ$\cdot$mol$^{-1}$K$^{-2}$. Therefore, the result suggests that the large $\gamma_{N}$ is related to the large DOS at the Fermi level as proposed in KNi$_{2}$Se$ _{2}$\cite{Fan2015}, instead of the heavy fermion state. For $A$Ni$_2$Se$_2$ ($A$ = K, Rb, and Cs), the increase of the atomic radiuses of alkali metals works as applying a pressure to the Ni-Se layers. The DOS of Ni $3d$ orbitals at the Fermi surface, the Sommerfeld coefficient $\gamma_{N}$, and the electronic correlations are all enhanced. In the framework of the BCS theory, the SC transition temperature $T_c$ increases accordingly, as the experimental observations in KNi$_2$Se$_2$, RbNi$_2$Se$_2$, and CsNi$_2$Se$_2$\cite{Neilson2012,Chen2016,Wang2013}.

\section{Summary}
	
In summary, we have successfully synthesized single crystals of RbNi$ _{2}$Se$_{2}$ and characterized the physical properties. RbNi$ _{2} $Se$ _{2} $ is found to be a weakly coupled superconductor with $T_c=1.2$ K. In normal state, RbNi$ _{2} $Se$ _{2} $ exhibits Fermi liquid behavior and Pauli paramagnetism. In the SC state, the zero-field electronic specific heat can be well described with a two-gap BCS model, indicating that RbNi$ _{2} $Se$ _{2} $ possesses multi-gap feature. DFT calculations and ARPES measurements demonstrate that multi electronic bands of the $3d$ orbitals of Ni ions cross the Fermi level. Our analyses reveal that the large Sommerfeld coefficient of RbNi$ _{2} $Se$ _{2}$ is originated from the large DOS at the Fermi surface.

\section{acknowledgments}

We thank Shiliang Li for fruitful discussions. Work at Sun Yat-Sen University was supported by the National Natural Science Foundation of China (Grants No. 12174454, 11904414, 11904416, 11974432, U2130101), the Guangdong Basic and Applied Basic Research Foundation (No. 2021B1515120015), National Key Research and Development Program of China (No. 2019YFA0705702, 2018YFA0306001, 2017YFA0206203), and GBABRF-2019A1515011337. Work at SLAB was supported by NSF of China with Grant No. 12004270, and GBABR-2019A1515110517. ARPES work at Rice is supported by the Robert A. Welch Foundation Grant No. C-2024 (M. Y.). P. D. is also supported by US Department of Energy, BES under Grant No. DE-SC0012311. Work at Berkeley was funded by the U.S. Department of Energy, Office of Science, Office of Basic Energy Sciences, Materials Sciences and Engineering Division under Contract No. DE-AC02-05-CH11231 (Quantum Materials program KC2202).

\bibliography{reference}

\end{document}